# Laser-assisted Oxidation of Multi-layer Tungsten Diselenide Nanosheets

C. Tan,[a,b] Y. Liu,[c] H. Chou,[b] J.-S. Kim,[a] D. Wu,[c] D. Akinwande,[a,*] and K. Lai[c,*]

a)  Microelectronics Research Center, University of Texas at Austin, Austin, TX 78758, USA

b)  Department of Mechanical Engineering and the Materials Science and Engineering Program, University of Texas at Austin, Austin, Texas 78712, USA

c)  Department of Physics, University of Texas at Austin, Austin, TX, 78712, USA

* E-mail: (K.L.) kejilai@physics.utexas.edu; (D.A.) deji@ece.utexas.edu.

## Abstract

We report the structural and electrical characterization of tungsten oxides formed by illuminating multi-layer tungsten diselenide ($WSe_2$) nanosheets with an intense laser beam in the ambient environment. A noninvasive microwave impedance microscope (MIM) was used to perform electrical imaging of the samples. The local conductivity $\sim 10^2$ S/m of the oxidized product, measured by the MIM and conventional transport experiments, is much higher than that of the pristine $WSe_2$, suggesting the formation of sub-stoichiometric $WO_{3-x}$ polycrystals with n-type carriers. With further efforts to improve the conductivity of the oxides, the laser-assisted oxidation process may be useful for patterning conductive features on $WSe_2$ or forming electrical contacts to various transition metal dichalcogenides.

Transition metal dichalcogenides (TMDCs) such as $MoS_2$ and $WSe_2$ are currently in the spotlight of scientific research due to their thickness-dependent semiconducting gaps and novel valley physics[1-4]. To fabricate TMDC-based nanoelectronic and optoelectronic devices, much effort has been made to understand the solid-state chemistry during the oxidation, etching, and metal deposition processes of the materials[5]. These standard semiconductor processing steps could strongly influence or even completely alter the electrical and optical properties of TMDCs. It is thus crucial to locally characterize the materials, preferably in a noninvasive manner, during the aforementioned fabrication procedures.

Laser-assisted processes are particularly important in the research of TMDCs and other two-dimensional (2D) materials. Raman spectroscopy, for instance, has been widely used to identify the number of layers from the frequency shift of specific vibrational modes[6,7]. Using the same Raman setup, it was shown that few-layer $MoS_2$ can be thinned down to monolayers under high laser intensity[8]. In addition to this ablation effect, laser-induced oxidation may also occur for graphene[9-11], phosphorene[12,13], and various TMDCs[14] under the ambient condition, while the electrical properties of the oxidized products have not been fully characterized. In this Letter, we report the structural and electrical studies of $WSe_2$ nanosheets subjected to intense illumination in air. Using microwave impedance microscopy (MIM) and transport measurements, we show that the oxidized regions, presumably forming sub-stoichiometric $WO_{3-x}$ with n-type carriers, are much more conductive than the pristine $WSe_2$. The laser-assisted oxidation thus provides a convenient way to create pre-defined conductive patterns, which may be useful for studying mesoscopic physics or forming electrical contacts on certain TMDC samples.

The $WSe_2$ nanosheets in this study were mechanically exfoliated onto 285 nm thick $SiO_2$ on heavily doped Si substrates. Since laser ablation of the sample dominates in few-layer TMDCs[8], we purposely chose multi-layer (10 – 100 nm) WSe$_2$ samples for this experiment. The relatively thick nanosheets ensured the efficient absorption of laser power for local heating and chemical reactions. The laser-assisted oxidation was performed in the WiTec Alpha 300 micro-Raman confocal microscope with a laser wavelength of 488 nm and a grating of 1800 lines/mm. As schematically shown in Fig. 1a, the laser beam was focused on the sample surface with a maximum intensity of 12 mW and a spot size of ~ 300 nm to induce oxidation under the ambient environment. The mapping resolution of the raster scan and integration time were set to 16 pixels/micron and 2 s, respectively.

Several microscopy and spectroscopy techniques have been employed to reveal the structural properties of the laser-oxidized products. Fig. 1b displays the Raman spectra acquired with a low laser power (<1 mW) at three locations of the sample shown in the inset. For the area that was not exposed to intense illumination, the prominent Raman peak at ~248 cm$^{-1}$ can be attributed to the E$^1_{2g}$ mode of pristine WSe$_2$ [15]. Several Raman features were observed at the oxidized part, and the dominant peak at ~805 cm$^{-1}$ is consistent with the value reported for crystalline WO$_3$ [16,17]. To obtain the elemental information, time-of-flight secondary ion mass spectroscopy (ToF-SIMS) was employed to analyze the two oxidized flakes in Fig. 1c, which showed strong WO$_2^-$ (216 atomic mass units, amu) signals in the ToF-SIMS data. Interestingly, clear Se$^-$ (80 amu) signals appeared at the bottom of the flakes, indicative of intact WSe$_2$ here. This observation was further corroborated by the cross-sectional transmission electron microscopy (TEM) image taken on a different sample, as depicted in Fig. 1d. While the 3D polycrystalline WO$_3$ structures can be seen in the bulk of the flake, van der Waals gaps of WSe$_2$ layers are visible just above the SiO$_2$/Si substrate. The result is reminiscent of the laser thinning process on TMDCs[8], where energy dissipation through the substrate was believed to preserve the bottom monolayer from being ablated. Note that the residual WSe$_2$ layers were only seen in relatively thick (> 50 nm) flakes, while thinner (< 20 nm) samples were usually fully oxidized under the above laser setting.

In order to study the electrical properties of the oxidized flakes, a microwave impedance microscope (MIM) based on the Park Systems XE-70 atomic-force microscopy (AFM) setup was used to perform non-destructive conductivity imaging experiment[18,19]. The MIM measures the real (MIM-Re) and imaginary (MIM-Im) parts of the tip-sample admittance (reciprocal of impedance), from which the local conductivity can be extracted with a spatial resolution comparable to the tip diameter[20]. Fig. 2a shows the optical picture and simultaneously acquired AFM and MIM images of a WSe$_2$ nanosheet, whose upper part was raster scanned by the intense laser prior to the MIM imaging. While the oxidation only resulted in a mild decrease of the thickness from 16 nm to 15 nm, both the optical contrast and MIM signals differ substantially between the pristine and oxidized sections of the sample, as is evident from the line profiles in Fig. 2b. For a semi-quantitative understanding of the MIM signals, finite-element analysis (FEA) was carried out to calculate the near-field microwave response[20], assuming a tip diameter of 300 nm and a dielectric constant of 20 for tungsten oxide[21-23]. As shown in Fig. 2c, the measured MIM contrast (~3.2 V in MIM-Im and ~1.5 V in MIM-Re) corresponds to a sheet resistance around $10^6$ Ω/sq or an effective conductivity on the order of $10^2$ S/m. The MIM contrast, thus the conductivity, is stable over weeks of storage under the ambient condition. We have measured over ten laser-oxidized WSe$_2$ nanosheets in the MIM experiments, all of which showing similar conductivity of the oxides. Since the stoichiometric WO$_3$ is a good insulator with a band gap of ~3 eV [21-23], it is likely that the high conductivity is associated with the formation of conductive sub-stoichiometric WO$_{3-x}$ [24] during the laser-assisted oxidation process.

To quantitatively extract the DC conductivity of $WO_{3-x}$ and compare with the MIM data, the sample in Fig. 2 was fabricated into a field-effect transistor for conventional transport measurements. Fig. 3a shows the $I_{DS}$-$V_{DS}$ characteristics at zero gate voltage of three adjacent segments across $WSe_2$-$WSe_2$, $WSe_2$-$WO_{3-x}$, and $WO_{3-x}$-$WO_{3-x}$ regions, respectively (color-coded in the inset). Ohmic contact between Pd electrodes and $WO_{3-x}$ was obtained, as compared to the Schottky behavior and very low source-drain current on the $WSe_2$ side. The sheet resistance of $WO_{3-x}$, ~ 3 MΩ/sq from four-point measurement (not shown), is in reasonable agreement with the MIM result. The gate dependence in Fig. 3b further demonstrates that, in contrast to the typical ambipolar behavior of the pristine $WSe_2$ [25, 26], the laser-oxidized $WO_{3-x}$ is a conductor with n-type carriers presumably due to the oxygen deficiency. The weak gate dependence is indicative of the low carrier mobility of $WO_{3-x}$ formed under the ambient environment.

Finally, we briefly compare the laser-assisted oxidation with the thermal oxidation of $WSe_2$ reported in Ref. [27]. In both cases, the MIM is a useful method to map out the spatial distribution of local conductivity. For the thermal oxidation at elevated temperatures, it was shown that the reaction starts from the edges and propagates toward the center for thick films[27,28], or results in triangular etch pits for atomically thin layers[29]. From the MIM data, the interface between $WSe_2$ in the interior and fully oxidized $WO_3$ in the perimeter also exhibits a low sheet resistance of $10^5 - 10^6$ Ω/sq [27], which again suggests the formation of sub-stoichiometric $WO_{3-x}$ here. On the other hand, the high-intensity laser beam can be manipulated to enable on-demand writing[8,30] of conductive features on the $WSe_2$ nano-sheet samples. With future efforts to enhance the carrier mobility in $WO_{3-x}$ and improve the $WO_{3-x}$-$WSe_2$ interface, such as thermal annealing in an inert atmosphere, it may be possible to take advantage of its high conductivity to form low-resistance electrical contacts to $WSe_2$ or other TMDCs.

In summary, we report the structural and electrical characterization of the oxidized $WSe_2$ nanosheets subjected to intense laser illumination. Similar to the thermal oxidation process, the laser-assisted reaction with oxygen in the ambient leads to the formation of sub-stoichiometric $WO_{3-x}$ polycrystals. Both microscopic electrical imaging and macroscopic transport measurements revealed the high local conductivity of the oxides, which may be utilized to create conductive patterns or to reduce contact resistance in $WSe_2$ nano-devices. In this regard, noninvasive electrical mapping by microwave microscopy could be of particular importance for future applications of TMDCs and other 2D materials.

The MIM work by Y.L. D.W., and K.L. was supported by Welch Foundation Grant F-1814. C.T. acknowledges support from a National Defense Science and Engineering Graduate (NDSEG) Fellowship: Contract FA9550-11-C-0028, awarded by the Department of Defense. Work by C.T., J.K., and D.A. was supported in part by the Southwest Academy of Nanoelectronics (SWAN), a Nanoelectronics Research Initiative (NRI) center.

**Figures:**

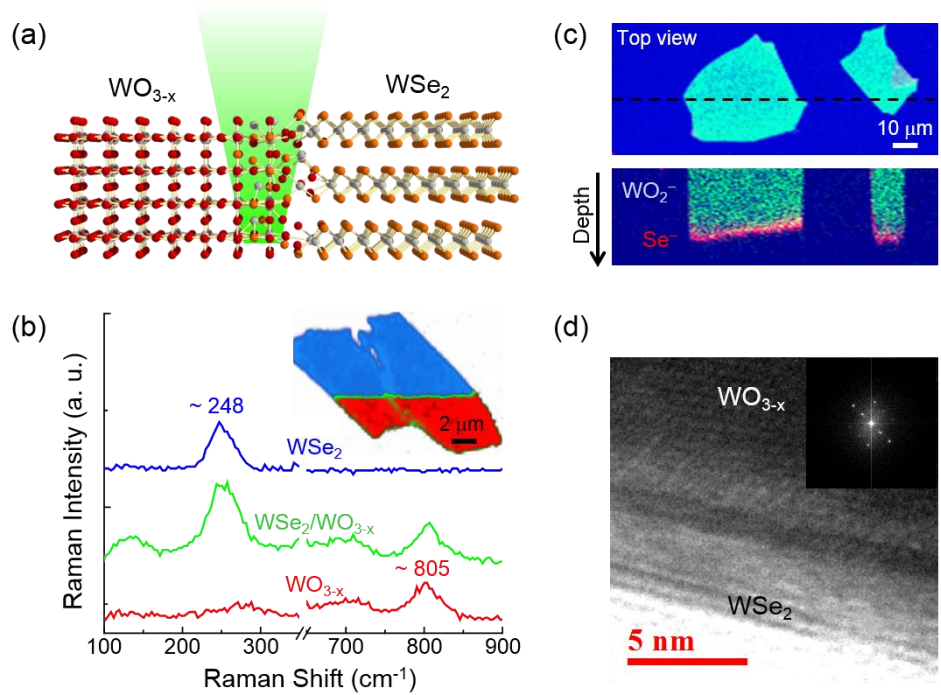

Figure 1: (Color online) (a) Schematic of the laser-assisted oxidation of the layered $WSe_2$ into the 3D $WO_3$. (b) Typical Raman shifts taken at the pristine $WSe_2$ (blue), the $WSe_2/WO_{3-x}$ interface (green), and the $WO_{3-x}$ regions (red), respectively. Characteristic Raman peaks of $WSe_2$ and $WO_3$ are labeled in the plot. The inset shows the half-oxidized sample, in which the Raman data were acquired. The sum of Raman intensity counts in the range of 220 – 270 cm$^{-1}$ is shown in blue and 770 – 840 cm$^{-1}$ in red. The green contours are the overlapping regions between the blue and the red areas. The scale bar is 2 μm. (c) (Top) ToF-SIMS map of two oxidized flakes and (bottom) depth profile of the Se$^-$ (red) and $WO_2^-$ (cyan) signals during the ion sputtering. The scale bar is 10 μm. (d) Cross sectional TEM image of an oxidized sample with its FFT image in the inset. Signatures of the layered $WSe_2$ can be seen at the bottom surface of the sample. The scale bar is 5 nm.

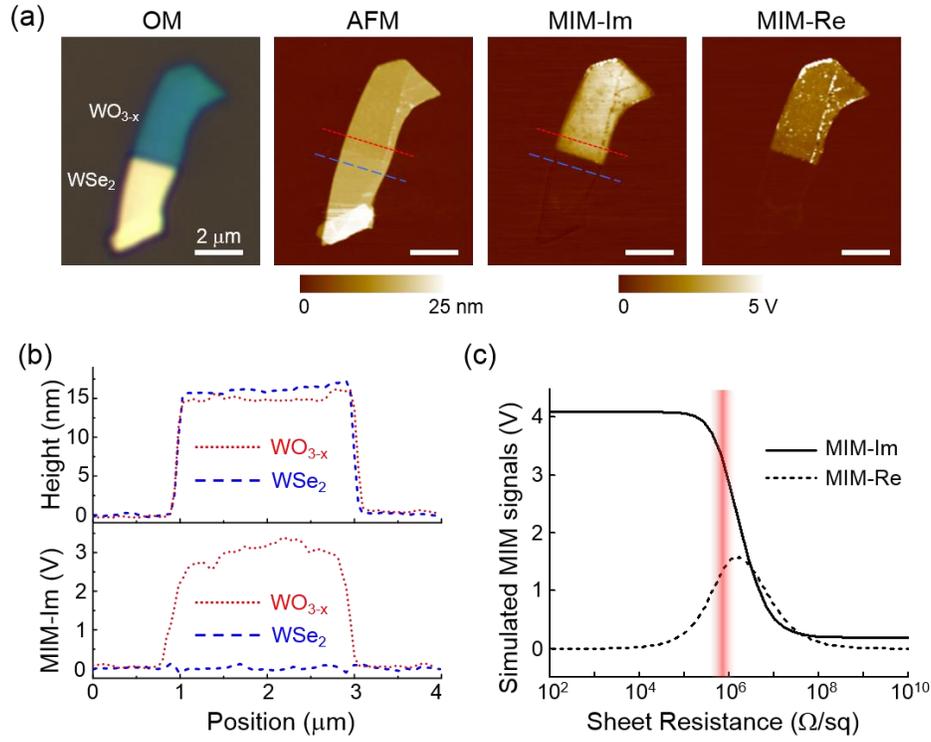

Figure 2: (Color online) (a) From left to right – Optical, AFM, and MIM-Im/Re images of a WSe$_2$ nano-flake whose upper portion was subjected to laser oxidation. All scale bars are 2 μm. (b) Line profiles of the (top) topography and (bottom) MIM-Im data indicated in (a). (c) Simulated MIM signals as a function of the sheet resistance $R_{sh}$ of the 15 nm WO$_{3-x}$. The measured MIM signals are shown as the shaded column, which corresponds to $R_{sh} \sim 10^6$ Ω/sq.

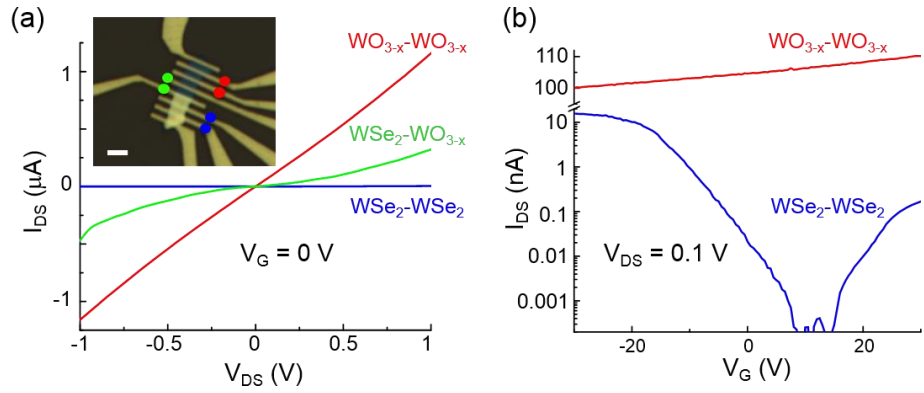

Figure 3: (Color online) (a) I-V characteristics across three pairs of Pd electrodes of the device shown in the inset. The scale bar is 2 μm. (b) Gate dependence on the ambipolar $WSe_2$ segment (blue) and the n-type $WO_{3-x}$ segment (red) of the sample.